\documentstyle[twocolumn,aps]{revtex}

\begin{document}

\title{POLYMERIC PHASE OF SIMPLICIAL QUANTUM GRAVITY}
\author{Davide Gabrielli} 

\address{Scuola Internazionale Superiore di Studi Avanzati}

\address{via Beirut 2-4 34013 Trieste, Italy}

\maketitle

\begin{abstract}
We deduce the appearance of a polymeric phase in 4-dimensional simplicial
quantum gravity by varying the values of the coupling constants and
discuss
the geometric structure of the phase in terms of ergodic moves. A similar
result is true in 3-dimensions.
\end{abstract}


\section{INTRODUCTION}
Path-integral approach to quantum gravity leads in a natural way to 
geometric and probabilistic problems that have a great interest  
both in physics and mathematics
\cite{1Ambjorn}, \cite{Mauro}. A basic issue of the theory
is to make sense of a formal probability measure on 
Riemannian structures associated with the partition function
\begin{equation}
{\cal Z}=\sum_{Top}\int_{\frac{Riem(M)}{Diff(M)}}{\cal D}[g(M)]e^{-S[g,M]}
\end{equation}
where the action $S$ is defined as a 
functional over Riemannian manifolds $(M,g)$ 
\begin{equation} 
S[M,g]=\Lambda \int_{(M,g)} dx^{4}\sqrt {g}- {1\over 16\pi G}
\int_{(M,g)} dx^{4}\sqrt {g} R
\label{action}
\end{equation}
with $\Lambda$ the cosmological constant and $G$ 
the gravitational constant.
${\cal Z}$ is a badly ill-defined quantity since both
the sum over 
topologies and the integration on the space of 
Riemannian structures ${\frac{Riem(M)}{Diff(M)}}$ cannot be given any
sensible mathematical status.

\noindent The idea of simplicial quantum gravity is 
the classical and often useful idea of bypassing such issues
by discretizing the theory and 
recovering the continuum one with a suitable limiting procedure.
Following Regge \cite{Regge} we use as discrete Riemannian 
manifolds piecewise flat
manifolds obtained by gluing together a finite number of simplices.
Briefly: inequivalent (in the sense of Tutte) dynamical triangulations
will simulate the inequivalent Riemannian structures; the vector
\begin{equation}
f=(N_0,N_1,N_2,N_3,N_4)
\label{vettore}
\end{equation}
with $N_i$ the number of $i$ dimensional simplices is called the $f$ 
vector of the triangulation and the curvature is concentrated
on the $d-2$ dimensional simplices (bones): the curvature
on the bone $b$ is
\begin{equation}
K(b)=\frac{48\sqrt{15}}{a^2}
\left[\frac{2\pi -q(b)cos^{-1}\frac{1}{4}}{q(b)}\right]
\end{equation}
where $q(b)$ is the number of $d$ dimensional simplices
incident on $b$ and $a$ is the length of the sides of the 
gluing simplices. The discrete counterpart of the action
(\ref{action}) is
\begin{equation}
S=k_4N_4-k_2N_2
\end{equation}
where $k_4$ depends linearly on the inverse 
of the gravitational constant and on the cosmological constant,
whereas $k_2$ is proportional to the inverse of the
gravitational constant. Restricting to triangulations with spherical
topology,
the partition function of the discrete theory is:
\begin{equation}
{\cal Z}(k_2,k_4)=\sum_{T\in S^4}e^{k_2N_2(T)-k_4N_4(T)}.
\end{equation}
This can be written more explicitly in the form
\begin{eqnarray}
{\cal Z}(k_2,k_4)&=&\sum_{N_4}\sum_{N_2}
W(N_2,N_4)e^{k_2N_2-k_4N_4} \nonumber \\
&=&\sum_{N_4}Z(N_4,k_2)e^{-k_4N_4}
\label{sistema}
\end{eqnarray} 
where $W(N_2,N_4)$ is the number of inequivalent spherical
triangulations with $N_4$ simplices and $N_2$ bones, and $Z(N_4,k_2)$ is
the canonical partition function (at fixed volume).
It is important to stress that $N_4$ and $N_2$ completely
determine the $f$ vector due to the Dhen-Sommerville relations:
\begin{equation}
\sum_{i=0}^{4}(-1)^{i}{N_{i}(T)}=\chi(T)   
\label{eupo} 
\end{equation}
\begin{equation}
\sum_{i=2k-1}^{4}(-1)^{i}{(i+1)! \over (i-2k+2)!(2k-1)!}N_{i}(T)=0  
\label{deso}  
\end{equation}
with $ k=1,2$ . Equation (\ref{eupo}) is just the
Euler-Poincar\'e equation while (\ref{deso}) are
consequence of the fact that the link
of every $2k$-simplex is an odd dimensional sphere, and hence has Euler number
zero.
Our interest will be concentrated on the statistical system (\ref{sistema}).

\section{GEOMETRICAL CONSTRAINTS AND ERGODIC MOVES}
\label{moves}

\noindent There exists a set of ergodic moves (elementary surgery operations)
in $3$ and $4$ dimensions called 
generically $(k,l)$ moves \cite{Gross}: $k$ and
$l$ are integers numbers such that $k+l=d+2$. The moves consist
in cutting out a subcomplex made up of $k$ simplices substituting
it with a complex of $l$ simplices with the same boundary. In particular,
if $s^d$ is the $d$-dimensional simplex, the $k$ complex is the 
star of a $d-k+1$ simplex in $\partial s^{d+1}$ and the $l$ complex is the
complement (see \cite{Gross} for a detailed description). In this way all
spherical triangulations can be constructed starting from the basic
$\partial s^{d+1}$ with a finite number of moves 
(actually in $4d$ this is true 
only for smooth triangulations). Following this construction we can
give a characterization of the generic $f$ vector by analyzing how 
$(k,l)$ moves modify it:
\begin{equation}
(1,5)\ \rightarrow \ \Delta_{1,5}f=(1,5,10,10,4)
\end{equation}
\begin{equation}
(2,4)\ \rightarrow \ \Delta_{2,4}f=(0,1,4,5,2)
\end{equation}
\begin{equation}
(3,3)\ \rightarrow \ \Delta_{3,3}f=(0,0,0,0,0)
\end{equation}
and obviously $\Delta_{5,1}f=-\Delta_{1,5}f$ and 
$\Delta_{4,2}f=-\Delta_{2,4}f$.
If $n_{k,l}$ is the number of moves of the type $(k,l)$ the 
corresponding $f$ vector will be
\begin{eqnarray}
f&=&(6+x_1,15+5x_1+x_2,20+10x_1+4x_2, \nonumber \\
& &15+10x_1+5x_2,6+4x_1+2x_2)
\end{eqnarray}
with $x_1=n_{1,5}-n_{5,1}$ and $x_2=n_{2,4}-n_{4,2}$.

\noindent This characterization of the $f$ vector is equivalent
to (\ref{vettore}) with the Dhen-Sommerville relations. This is not enough
to completely determine the possible $f$ vectors since
 it is not always possible to perform a $(k,l)$ move.
It is always possible apply a move of the type $(1,5)$
but to apply the reverse move we must start from a triangulation
that has a vertex with a star
made of $5$ simplices. It is often
possible to
apply a move of type $(2,4)$ but in order
to apply the reverse move we
must start from a triangulation that has 
an edge with a star made of $4$ simplices.
\noindent A result in this direction is the following, essentially due
to Walkup \cite{Walkup}:

\noindent {\bf Theorem} {\it
For any combinatorial triangulation of a $4$ sphere the inequality
\begin{equation}
N_1\geq 5N_0-15
\end{equation}
holds with equality if and only if it is a stacked sphere.}

\noindent A $d$-dimensional stacked sphere is a triangulation obtained
from $\partial s^{d+1}$ applying only $(1,d+1)$ moves. 

\noindent To this lower bound theorems we can add a more trivial upper bound
\begin{equation}
N_1\leq\frac{N_0(N_0-1)}{2}
\label{banana}
\end{equation}
that only says that evidently the edges must be less than all the possible
couple of vertices, together with the obvious condition 
\begin{equation}
N_0\geq d+2
\end{equation}

\noindent Translating this inequalities in terms of moves we obtain
\begin{equation}
x_2\geq 0,\ \  x_1\geq 0,\ \ x_1^2+x_1-2x_2\geq 0
\label{vero3}
\end{equation}
It is interesting also to write
explicitly these constraints in terms of the parameters that are usually
used, $N_2$ and $N_4$, since
we will be interested in the statistic behavior at fixed volume:
\begin{equation}
N_2\leq \frac{5}{2}N_4+5,\ \ N_2\geq 2N_4+8
\label{prima}
\end{equation}
\begin{equation}
9N_2^2-18N_2-12N_2N_4+24N_4+4N_4^2\geq 0
\label{seconda}
\end{equation}
It will be very useful \cite{Mauro} in 
the study of the asymptotic behavior
of our model to introduce the parameter
$\xi=\frac{N_2}{N_4}$. Also the 
geometrical constraints become more easy, in the large volume limit, using this
parameter; equations (\ref{prima}) tell to us that
\begin{equation}
\xi\leq\frac{5}{2}+\frac{5}{N_4},\ \ \xi\geq2+\frac{8}{N_4}
\end{equation}
that in the limit $N_4\rightarrow \infty$ become
\begin{equation}
2\leq\xi\leq\frac{5}{2}
\label{limitato}
\end{equation}
From equation (\ref{seconda}) we obtain
\begin{equation}
9\xi^2-18\frac{\xi}{N_4}-12\xi +\frac{24}{N_4}+4\geq 0
\end{equation}
and using the bounds (\ref{limitato}) we obtain
\begin{equation}
9\xi^2-12\xi+4=(3\xi-2)^2\geq 0
\end{equation}
that is always true.

\noindent These asymptotic conditions are consequences 
of Walkups theorem whose demonstration is in fact quite
not trivial. However, we can
give a simple argument in terms of moves providing an intuitive
picture: as we have already stressed it is 
far more easy to perform
a $(k,l)$ move with $k<l$ than the reverse one and such a move
increases the volume of the manifold while the reverse move 
decreases it;
so we can conclude that, 
when the number of simplices is large, almost all (in fact Walkups
theorem say all 
(\ref{vero3})) triangulations are obtained with a number
of $(k,l)$ moves greater than $(l,k)$. So we can obtain the conditions
(\ref{limitato}) as limiting values in the
boundaries of the allowed region:
\begin{equation}
\xi_{min}=\lim_{x_2\rightarrow \infty}\xi 
=\lim_{x_2\rightarrow \infty}\frac{20+10x_1+4x_2}{6+4x_1+2x_2}=2
\end{equation}
\begin{equation}
\xi_{max}=\lim_{x_1\rightarrow \infty}\xi 
=\lim_{x_1\rightarrow
\infty}\frac{20+10x_1+4x_2}{6+4x_1+2x_2}=\frac{5}{2}  
\end{equation}
A funny dynamical interpretation of the constraints can
be given in terms of equilibrium points of the ``moves operators'':
starting from a triangulation with $\xi=\frac{N_2}{N_4}$ we have
a jump
\begin{equation}
\Delta_{1,5}\xi=\frac{N_2+10}{N_4+4}-\frac{N_2}{N_4}
\end{equation}
The equilibrium condition is
\begin{equation}
\Delta_{1,5}\xi=0 \leftrightarrow \xi=\frac{5}{2}
\end{equation}
and this equilibrium point is stable in the sense that
\begin{equation}
\Delta_{1,5}\xi >0 \leftrightarrow \xi <\frac{5}{2},\ \  
\Delta_{1,5}\xi <0\leftrightarrow \xi >\frac{5}{2}
\label{stabil1}
\end{equation}
Likewise for move $(2,4)$
\begin{equation}
\Delta_{2,4}\xi=0 \leftrightarrow \xi=2
\end{equation}
and also this equilibrium point is stable in the sense (\ref{stabil1}).
This simple analysis explain why only the region
(\ref{limitato}) is spanned when 
constructing
spherical triangulations with a large number of simplexes: because
points $\xi$ that are outside this interval are 
attracted towards it.

\section{ASYMPTOTIC BEHAVIOR OF CANONICAL MEASURE}
\label{as}
The behavior of the system conditioned to fixed volume is described
by canonical partition function
\begin{equation}
Z(k_2,N_4)=\sum_{T_{N_4}}e^{k_2N_2} 
=\sum_{N_2}W(N_2,N_4)e^{k_2N_2}
\label{partit}
\end{equation}
We can study it by using the parameter $\xi$ \cite{Mauro}:
\begin{equation}
Z(k_2,N_4)=\sum_{k}W(N_4,\xi_{k}N_4)e^{k_2N_4\xi_{k}}
\end{equation}
Since the number of triangulations with $N_4$ simplexes is 
asymptotically exponentially bounded (see \cite{Mauro} 
for a demonstration), 
the asymptotic behavior of $W(N_4,\xi_{k}N_4)$ 
can be formalized in the form
\begin{equation}
W(N_4,\xi_kN_4)\sim f(N_4,\xi_k) e^{N_4s(\xi_k)}
\end{equation} 
with $f(N_4,\xi)$ that has typically a polynomial 
or subexponential asymptotic behavior in $N_4$. The measure
induced in this way in the space of triangulations is defined by the
probabilities
\begin{equation}
\mu_{k_2,N_4}^C(\xi_k)=\frac{f(N_4,\xi_k) e^{N_4(s(\xi_k)-k_2\xi_k)}} 
{Z(k_2,N_4)}
\end{equation}
The asymptotic behavior of probability measures defined in this
way is a classical problem of probability theory and under general
conditions the result is
\begin{equation}
\mu_{k_2,N_4}^C\Rightarrow_{N_4\rightarrow \infty}
\sum_{i}\mu_i \delta (\xi-\xi_{i}^*)
\label{convergenza}
\end{equation}
The points $\xi_{i}^*$ are defined by the condition
\begin{equation}
s(\xi_i^*)-k_2\xi_i^*=\sup_{\xi_{min}\leq\xi\leq\xi_{max}}[s(\xi)-k_2\xi]
\end{equation}
and the convergence (\ref{convergenza}) is very fast; namely considered
a set $A$ such that $\xi_i^*\notin A$ we have
\begin{equation}
\mu_{k_2,N_4}^C(A)\sim \frac{e^{N_4(\sup_{\xi\in A}[s(\xi)-k_2\xi])}}
{e^{N_4(s(\xi^*)-k_2\xi^*)}}
\sim e^{-KN_4}
\end{equation}
with $K >0$; that says that the probability of deviant events $A$
goes to zero exponentially fast: this fact is usually referred as the
deviations are large.

\noindent This general argument can be formalized in this particular
case in terms of Laplaces method: the form of the partition function
\begin{equation}
Z(k_2,N_4)=\sum_{k} f(N_4,\xi_k) e^{N_4(s(\xi_k)+k_2\xi_{k})}
\end{equation}
has the structure of a Riemann sum
\begin{equation}
Z(k_2,N_4)\sim\sum_{k}N_4 f(N_4,\xi_k) 
e^{N_4(s(\xi_k)+k_2\xi_{k})}\Delta(\xi_k)
\label{riemann}
\end{equation}
namely 
$\xi_{k+1}-\xi_{k}\sim \frac{1}{N_4}$ and we have a sum of 
$\frac{\xi_{max}-\xi_{min}}{\Delta\xi}\sim N_4$ terms. So for large $N_4$
(\ref{riemann}) is well approximated by the continuum version (see \cite{Mauro}
for details and an explicit form of $s(\xi)$ and $\tilde{f}(N_4,\xi)$ )
\begin{equation}
Z(k_2,N_4)\sim\int_{\xi_{min}}^{\xi_{max}} \tilde{f}(N_4,\xi) 
e^{N_4(s(\xi)+k_2\xi)} d\xi 
\label{integ}
\end{equation}
Laplaces theorem says that when $N_4$ is large almost all the 
contribution to the value of the integral (\ref{integ}) comes from the
region near the point(s) $\xi^*$; and this is a result of type 
(\ref{convergenza}).

\section{POLYMERIC PHASE}
We are now interested in a theoretic interpretation of numerical results
that give a strong evidence of the appearance of a polymeric phase
for $k_2$ large enough \cite{Ambjorn}. 
We will show that this phenomena is a direct
consequence of the concentration of the measure illustrated in the
previous chapter and we will analyze the geometrical characteristic
of this phase.

\noindent When $k_2$ is large, triangulations with 
large $\xi$ are favorite; 
we translate this simple idea into  a mathematical language:  
the points of maximum in the exponent of the expression (\ref{integ})
are determined by the condition
\begin{equation}
s'(\xi)+k_2=0
\label{deriva}
\end{equation}
from this it is immediately to deduce that if $k_2 > 
-\inf_{\xi_{min}\leq\xi\leq\xi_{max}}s'(\xi)$ the expression
(\ref{deriva}) is always greater than zero and the canonical measure
concentrates on $\xi_{max}$:
\begin{equation}
\mu_{k_2,N_4}^C(\xi)\Rightarrow_{N_4\rightarrow\infty}\delta (\xi-\xi_{max})
\end{equation}
These triangulations are well described by the lower bound 
result of Walkup
and we can conclude that in this region of the parameter $k_2$ the 
dominant configurations in the statistical sum (\ref{partit}) are 
essentially stacked spheres. A more precise statement could be that 
for the dominant configurations the most important phenomena is the
stacking ($(1,d+1)$ moves); a general characterization 
can in fact be given: starting from $y_1=\frac{N_4-2y_2-6}{4}$
we obtain 
\begin{equation}
\xi=\frac{5}{2}-\frac{y_2}{N_4}+\frac{5}{N_4}
\end{equation}
We can easily conclude that triangulations with large $N_4$ 
characterized
by a $\xi=\xi_{max}=\frac{5}{2}$ are obtained with the condition
\begin{equation}
\lim_{N_4\rightarrow\infty}\frac{y_2}{N_4}=0
\end{equation}
This condition is satisfied not only by stacked spheres, which are defined
by the relation $y_2=0$, but also by triangulations with $y_2=C<\infty$
that can be constructed, for example, by stacking starting from a 
generic triangulation
(defined by $y_2=C$) and not from the basic triangulation $\partial s^5$;
and more in general the condition is satisfied also by triangulations
constructed with a number of moves $y_2$ that grows with a 
power in $N_4$
smaller than one. 

\noindent The practical consequence of this assertions is that we 
can study
the statistical property of simplicial quantum gravity for that region
of $k_2$ by studying a more simple model obtained by restricting the 
space of configurations: the smaller statistical system that we will
consider is the system constituted by only stacked spheres. We have 
just showed that this in fact is a further simplification but
what happens is
that this subsystem contains all the principal features of interest.
In general the following trivial relation is true
\begin{equation}
Z(k_2,N_4)> \sum_{(S.S.)_{N_4}}e^{k_2N_2}
\end{equation}
but for $k_2$ large enough this
relation becomes
\begin{equation}
Z(k_2,N_4)\gtrsim\sum_{(S.S.)_{N_4}}e^{k_2N_2}
\label{sempart}
\end{equation}
where the abbreviation $(S.S.)$ means obviously (Stacked Spheres) and 
the symbol $\gtrsim$ means that the exact relation is $\geq$ but the
asymptotic behavior is the same $\sim$.

\noindent The study of the partition function for the 
stacked spheres system (\ref{sempart})
is an easier problem: the condition for stacked spheres 
$y_2=0$ tells
to us that $N_4=6+4y_1$ and $N_2=20+10y_1$ and we obtain the 
relation $N_2=\frac{5}{2}N_4+5$. This latter tells to us 
that the number of
bones is determined by the number of simplices. Consequently, 
the explicit 
expression of the partition function is:
\begin{equation}
Z_{S.S.}(k_2,N_4)=W_{S.S.}(N_4)e^{k_2(\frac{5}{2}N_4+5)}
\label{esseaesse}
\end{equation}
The problem that remain to face is the calculation of the number
of inequivalent stacked spheres with $N_4$ simplices: we will do it in
an approximate way stressing above all the fact that the structure of the
different configurations is typical of branched polymers with a fundamental
element (monomer) that builds up a tree configuration, and
we will also calculate the entropy exponent $\gamma$ and show that
it is $\frac{1}{2}$ as is typical of branched polymers.

\noindent The tree structure that is behind stacked spheres can be 
easily reconstructed from the following geometrical 
interpretation of $(1,5)$ moves: starting from $s^5$ we construct the
basic spherical $4$-d triangulation as $\partial s^5$; a $(1,5)$ move
is obtained by substituting a $4$-d simplex of this triangulation with
5 simplices as discussed in section (\ref{moves}); but we can proceed also
in a different way by constructing a new triangulation of the $5$-d ball
gluing a second $s^5$ trough a $4$-d face to the beginning simplex. The 
spherical triangulation $\partial B$ where $B$ is the new triangulating
ball obtained in this way is equivalent to the triangulation
obtained with the $(1,5)$ move. This construction is general: we obtain 
every stacked sphere as the boundary $\partial B_{n}$ of triangulations of
the $5$-d ball obtained by gluing a $s^5$ to $B_{n-1}$ through a $4$-d face
of $\partial B_{n-1}$ and this correspondence is easily seen to be
one to one \cite{Walkup}. 
It is also easily to see that $\partial B_{n}$ is a stacked
sphere with $2+4n$ simplices. This construction is illuminating in 
characterizing a polymeric phase 
in simplicial quantum gravity. The monomer with which
the polymer is builded is provided by the  $s^5$ simplices
and the polymer structure
is obtained by analyzing the only tree-like triangulations of $5$-d ball 
whose boundary are stacked spheres. A good insight in the structure of such
polymers and also a tool for calculating $W_{S.S.}(N_4)$ is obtained by
analyzing the $1$-d skeleton of the dual of $B_{n}s$. To every $s^5$ it is 
associated a point and in each such points there are 6 lines incident
that correspond to the 6 $4$-d faces; when two $s^5$ are glued along
a face the corresponding points are joined by a line. The graphs 
obtained in this way are all the possible trees the incidence numbers
of which are only $6$ and $1$: the vertices with incidence 
$6$ represent the 
$5$-d simplices glued together and the vertices with incidence $1$
represents the free $4$-d faces of $\partial B_{n}$. This construction
is not enough to reconstruct the ball $B_{n}$, and the corresponding
stacked sphere. This reconstruction would be 
possible only from the knowledge
of the full dual structure, nonetheless the above partial construction
will be enough to calculate the right
asymptotic behavior. 

\noindent The problem of counting such trees is equivalent to a problem
of counting isomers in chemistry: a solution is given in the classical
paper of Otter \cite{Otter}. The asymptotic expression of the number
of not isomorphic trees with $n$ vertex and ramification 
number not greater than $m$ is:
\begin{equation}
T_n^{<m}\sim c(m)\frac{\alpha(m)^{n}}{n^{\frac{5}{2}}}
\end{equation}
This is easily seen to be also the solution of our counting problem:
namely the following relation holds
\begin{equation}
T_n^{<m}=T^{(1,m)}_{((m-2)n+2,n)}
\end{equation}
The notation of the left side was already explained and the symbol
on the right side means the number of trees with  $(m-2)n+2$ vertices 
with number of incidence $1$ and $n$ vertices with
number of incidence $m$.
The equality is verified by constructing explicitly a one to one 
correspondence: starting from a tree in $T^{(1,m)}_{((m-2)n+2,n)}$
we delete all vertices with number of incidence $1$ and the corresponding
lines and we obtain an element in $T_n^{<m}$; the reverse correspondence
is obtained by joining new vertices with ramification number $1$ to
the old vertices until all the old vertices reach the exact ramification
number $m$.

\noindent In order to count 
the number of stacked spheres with $N_4$ simplexes
we have to estimate the number of inequivalent trees with $\frac{N_4-2}{4}$
vertices with incidence $6$ and $N_4$ vertices 
with incidence $1$; we get
\begin{eqnarray}
W_{S.S.}(N_4)&\gtrsim & T^{(1,6)}_{(N_4,\frac{N_4-2}{4})}
=T^{<6}_{\frac{N_4-2}{4}} \\ \nonumber 
&=&c\frac{\alpha^{\frac{N_4-2}{4}}}{\left(\frac{N_4-2}{4}\right)
^{\frac{5}{2}}}\sim cost\frac{\left(\alpha^{\frac{1}{4}}\right)^{N_4}}
{N_4^{\frac{5}{2}}}
\end{eqnarray}

\noindent With this rough but effective asymptotic estimate we get 
informations about the canonical partition function by using relations
(\ref{sempart}), (\ref{esseaesse})
\begin{equation}
Z(k_2,N_4)\gtrsim C(k_2)\frac{1}{N_4^{\frac{5}{2}}}e^{N_4(\frac{1}{4}log\alpha 
+\frac{5}{2}k_2)}
\label{asin}
\end{equation}

Thus, we have obtained an expression of the form
\begin{equation}
f_{k_2}(N_4)e^{N_4k_4^c(k_2)}
\end{equation}
with a subleading asymptotics $f_{k_2}(N_4)$ of polynomial type. The 
subleading asymptotics is particularly important because from it we 
can deduce the entropy exponent $\gamma$: the general form is
$f_{k_2}(N_4)\sim N_{4}^{\gamma-3}$ that in our case gives 
$\gamma -3=-\frac{5}{2}$
and we obtain $\gamma=\frac{1}{2}$ as is typical of 
branched polymers and as comes out from numerical simulations \cite{Ambjorn}.
From the knowledge of this exponent we can for example obtain the critical
behavior of susceptibility \cite{Ambjorn}
\begin{eqnarray}
\sum_{r}G(r,k_2,k_4)&=&\frac{d^2}{dk_4^2}{\cal Z}(k_2,k_4) \nonumber \\
\sim (k_4-k_4^c)^{-\gamma}&=&(k_4-k_4^c)^{-\frac{1}{2}}
\end{eqnarray}
where $G(r,k_2,k_4)$ is the  
correlation function \cite{Ambjorn}.

\noindent From expression (\ref{asin}) we can also get an
estimate of the critical line that we expect to be quit good when
the parameter $k_2$ is large enough:
\begin{equation}
k_4^c(k_2)\gtrsim \frac{1}{4}log\alpha+\frac{5}{2}k_2
\end{equation}
This turn out to be in fact compatible both with numerical and analytical 
\cite{Mauro} results.

\section{CONCLUSIONS}
There exists a Walkups theorem also for the 3-dimensional case
\cite{Kuhnel}, \cite{Walkup}, and all the previous analysis can be
repeated. 

\noindent The asymptotic behavior of
canonical measure suggested in
section (\ref{as}) stresses the peculiar character of simplicial quantum 
gravity as a critical system: the measure concentrates on 
different regions of the space of configurations 
for different values of $k_2$. This characteristics allows, for example,
to compute the mean value of geometrical objects restricting on a smaller
region of the configuration space.
\begin{equation}
E_{\mu^C_{k_2,N_4}}(f)\Rightarrow_{N_4\rightarrow\infty}
E_{\delta(\xi-\xi^*(k_2))}(f)
\end{equation}
This is exactly the procedure followed to describe 
the structure of the polymeric phase and more informations could be obtained
with a detailed study of statistical mechanics of stacked spheres 
(correlations functions, for example).

\noindent A further step must be the comprehension of the geometric
structure of the crumpled phase. This is connected with the discovery
of upper bound theorems that substitute the trivial bound (\ref{banana});
recent results (see \cite{Catterall}, for example) suggest that the 
region of the phase space
that corresponds to the crumpled phase is characterized by the
appearance of singular structures.

\section{ACKNOWLEDGMENTS}
I thank J. Ambj{\o}rn, M. Carfora and G. Gionti for comments and
discussions and M. Carfora for pointing out the basic reference
\cite{Walkup}.

\end{document}